\pdfoutput=1

\documentclass[11pt]{article}

\usepackage[final]{acl}

\usepackage{times}
\usepackage{latexsym}

\usepackage[T1]{fontenc}

\usepackage[utf8]{inputenc}

\usepackage{microtype}

\usepackage{inconsolata}

\usepackage{graphicx}

\usepackage{graphicx}
\usepackage{float}
\usepackage{lipsum}
\usepackage{subcaption}

\title{Traits of a Leader: User Influence Level Prediction through Sociolinguistic Modeling}

\author{Denys Katerenchuk$^1$ \\
  $^1$The Graduate Center, CUNY\\
Computer Science\\
New York, USA \\
\texttt{dkaterenchuk@gradcenter.cuny.edu} \\
\And
Rivka Levitan$^{1,2}$ \\
$^2$Brooklyn College, CUNY\\
Computer and Information Science\\
Brooklyn, USA\\
\texttt{levitan@sci.brooklyn.cuny.edu} \\}

\begin{document}
\maketitle
\begin{abstract}
Recognition of a user's influence level has attracted much attention as human interactions move online.
Influential users have the ability to sway others' opinions to achieve some goals. As a result, predicting users' level of influence can help to understand social networks, forecast trends, prevent misinformation, etc.
However, predicting user influence is a challenging problem because the concept of influence is specific to a situation or a domain, and user communications are limited to text.
In this work, we define user influence level as a function of community endorsement and develop a model that significantly outperforms the baseline by leveraging demographic and personality data. This approach consistently improves RankDCG scores across eight different domains.   
\end{abstract}

\section{Introduction}

Influential users have the ability to influence others' behavior to achieve their own agenda. This agenda can be anything from an attempt to persuade a person to make a particular purchase to an attempt to overthrow the government. 
For example, consider the January 6th, 2021, events when a group of people stormed the US Capitol building. According to the New York Times, the group self-organized on websites such as Reddit\footnote{https://www.nytimes.com/2022/01/13/us/politics/jan-6-tech-subpoenas.html}. We believe that by assessing the ideas and the stance of the group leaders, it is possible to predict the severity of the situation and prevent such events from happening.

User influence prediction is a difficult problem for many reasons. First, the concept of user influence depends on a problem and its domain. In the literature, terms such as influencers, community endorsed person, community leader, opinion leader, and others correspond to some form of user leadership \cite{razis2020modeling}. Each term has its own definition with respect to the domain. As a result, the research in one domain might not work in a different domain. Secondly, user communications are limited to text, and textual comments provide limited information, making predicting the level of a user's influence difficult.

The main goal of our work is to predict a user's influence level from a single comment of at least 32 tokens. First, we define user influence level as a function of community endorsement based on users' comments and rewards. As a result, each user receives a k-index score, which defines the user level of influence in a particular discussion (Sec. \ref{sec:data}). 
Second, we create a strong baseline to predict a user influence level. 
Third, we leverage earlier research from social sciences to create a user-centric model. In particular, we introduce supplementary sub-tasks for user demographic and personality detection (Sec. \ref{sec:methods}). These tasks are combined in a single multi-task model to improve the latent state representation and the user influence level prediction. This paper shows that leveraging user-centric information improves influence level prediction across eight domains (Sec. \ref{sec:results}). 

\section{Related Work}
\label{sec:related_work}

The definition of an influential person is different with respect to a particular problem, and consequently, it is difficult to compare the results of previous studies. For this reason, we explore the relationships between users and influence in a broader spectrum. In particular, to find a correlation between user traits and influence, we review studies on behavior analysis in a corporate environment, prediction of popular content, and user leadership in general.  

The researchers tried to find an answer to what makes a person influential. Some early work in social sciences has found a correlation between influence and personality traits. For example, \citet{gehring2007applying} investigates the correlation between Meyers-Briggs type indicators (MBTI) \citep{myers1987description} and influence in a business environment. The work points out that 7 out of the 16 MBTI personality types are defined with words associated with influence. The team surveys 53 top managers giving them MBTI tests to validate this observation and show that 93\% of responders fall into one of the seven personality types. 
\citet{wang2015understanding} discover that extroverts are more likely to use phrases such as "so proud," "so excited," or "can't wait," which are positive and can affect their online social status. The fact that users use positive language to influence discussion is shown in the study on power relations detection on Wikipedia talk pages by \citet{danescu2012echoes}. Human language provides a rich source of information about a person's emotional state. 

One of the early prominent studies on predicting influential individuals from text is done by \citet{gilbert2012phrases}. This work predicts whether an email was written to somebody of a lower or higher status defined by the job title. E. Gilbert shows that predicting whether an email is written to someone of higher status is possible with an accuracy of 70.7\% and that language use differs among colleagues with different statuses. This research is based on an n-gram language model and a support vector machine model. The paper lists ranked phrases that contribute the most to one of the two-class predictions. For example, the top 3 phrases with the most weight in an email written to someone of lower status are: "Have you been," "You gave," and "We are in." Looking closer at the phrases, we notice that the first and the third phrases are also used as hedge phrases. These hedges are linguistic devices that indicate uncertainty and are commonly used to mitigate orders \cite{lakoff1975hedges}. \citet{rosenthal2014detecting} shows that other factors, such as age and gender, are related to influence. Belonging to the same gender or age group can be interpreted as characteristics of a concept known as social proof. Social proof is related to the fact that people are more likely to be influenced by someone similar to them. While many factors make someone influential, age, gender, hedges, and personality correlate with influence.

Several papers consider the task of detecting some form of influence. \citet{jaech2015talking} addresses the problem of predicting influential comments with the most karma points on Reddit\footnote{www.reddit.com}. The problem is constrained to ten comments posted around the same time.
This work asks which features, such as user reputation, graph, timing, lexicon, etc., contribute to comment score prediction. They discovered that user reputation does not significantly affect comment popularity except for the AskScience subreddit, where most influential users write almost 10\% of high-ranked comments. Also, they demonstrate that graph structure and timing features play a significant role in top-comment prediction. Leveraging these findings, \citet{zayats2018conversation} propose LSTM-based methods that embed the entire conversation thread structure. 
In other words, the model is trained to learn lexical and graph-structural features. This tree-LSTM model outperformed the text-based LSTM model. Neural networks have shown to be effective in predicting user influence \cite{razis2020modeling}. In particular, attention-based models \cite{vaswani2017attention} have shown to be successful in predicting post popularity from a title \citep{weissburg2021judging}. The growing popularity of graph neural networks \cite{wu2020comprehensive} has benefited the problem of predicting social influence. \cite{qiu2018deepinf} successfully apply graph neural networks that take into account local user network and lexical information. This approach unites both semantic features and graph structure in a single model.

After reviewing relevant literature, we notice a gap between the early research in understanding user behavior and recent works that mainly rely on advances in neural networks. With the development of neural networks, the focus on user-centric understanding has faded away. In this work, we leverage user-centric information to improve latent representation and, as a result, achieve high scores in predicting user status from a single comment. Graph neural networks have been shown to be successful in social network analysis. However, this work is constrained to text for two reasons: 1) unlike textual data, the graph structure is not always available for analysis (closed social network, dyadic conversations), and 2) improvements in the text-based model can be used in graph neural networks in future research.

\section{Data and Evaluation}
\label{sec:data}

\subsection{Reddit Data}
\label{sec:reddit}

Reddit is a discussion platform where users discuss any topics. The website is divided into subreddits, which are sub-communities with specific discussions. Reddit users can create posts to initialize a discussion or write comments in response. Each comment can earn or lose karma points. This karma score can be used as a proxy for a reward or community endorsement, making Reddit a good data source to study influence. 
This paper uses the Reddit dataset proposed by \citet{jaech2015talking,fang2016learning}. This dataset provides Reddit data collected between January 1, 2014, and January 31, 2015, from 8 subreddits: AskMen, AskScience, AskWomen, Atheism, ChangeMyView, Fitness, Politics, and WorldNews.  The dataset is a collection of posts and comments with additional information such as author, author flair, karma scores, etc. Our work is constrained to textual comments, and we do not use user or structure-related information. The comments' length is enforced to a minimum of 32 tokens (white space tokenization) and a maximum of 256 tokens. We sample 100k comments from each subreddit, with the AskScience subreddit containing only 33k comments. This data is further split into dev and test subsets 10\% each. 

\begin{figure}
\centering
\includegraphics[width=1.00\linewidth]{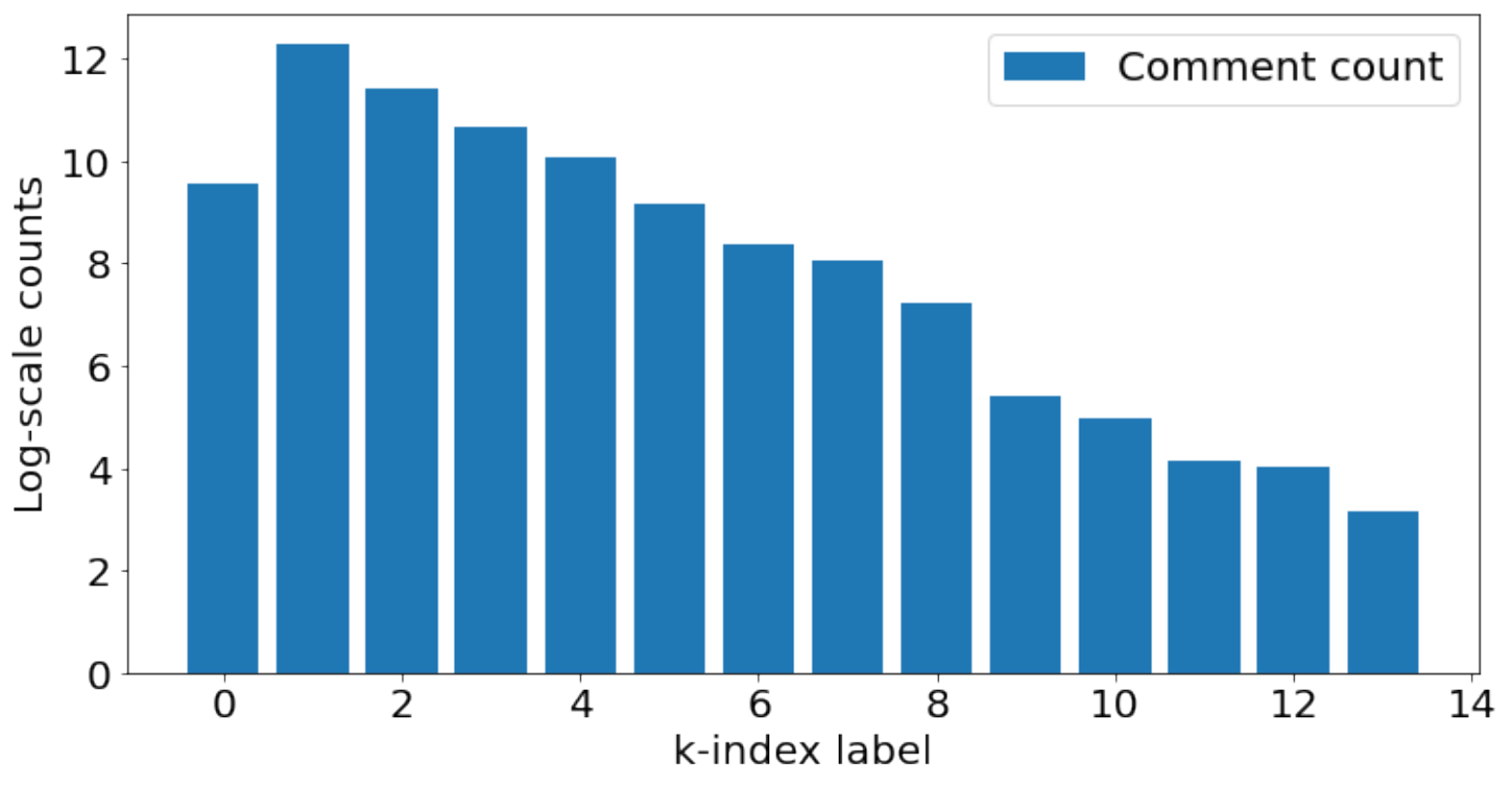}
\caption{Log-scale k-index distribution of AskScience subreddit.}
\label{fig:kindex_dist}
\end{figure}

This work adopts a k-index score to represent user status, which was introduced by \cite{jaech2015talking}. The k-index is defined as a maximum number of comments, let's say k, that has at least k karma. This score is essentially a modified h-index, an author-level metric \cite{hirsch2005index}. This score is used to mitigate outliers, where some comments can gain high karma scores for being one-off popular comments, off-topic, funny, etc. The k-index is calculated for each user in a single discussion thread and mapped to the corresponding author-written comments. We assume local user popularity, where a user might have a high status in one discussion thread but not another. We only consider the first 50 comments per discussion, disregarding the rest and discussion threads with less than 50 comments. 
One interesting characteristic of this data is that the k-index distribution is highly right-skewed, as seen in Figure \ref{fig:kindex_dist}. The figure shows log-scale k-index counts where most users have a k-index of 1, and very few have a k-index of 13. Hence, identifying rare, high-rank users is quite difficult. 


\subsection{Evaluation}
This work aims to develop new methods to improve user status prediction from a single comment. Considering the specifics of highly right-skewed k-index distribution where most users have a low k-index, and very few users have a high k-index, the task is defined as a ranking problem emphasizing identifying high-status users. 
For this reason, we use four evaluations: mean absolute error (MAE), mean squared error (MSE), normalized discounted cumulative gain (nDCG), and RankDCG. MAE and MSE are popular measures that help evaluate regression-based models by calculating the error between true and predicted labels. MAE reports the mean of the differences, and MSE is the mean of squares of the differences, which better represents high-status user prediction. NDCG \cite{jarvelin2002cumulated} is designed for information retrieval tasks. One advantage of nDCG measures is that it puts more emphasis on identifying high-status users. 
However, this measure has two noted downsides. First of all, this evaluation metric was designed for information retrieval rather than ordering evaluation. 
Secondly, the reward function depends on each element's position rather than relative class. 
All these shortcomings are explained and resolved in work by \citet{katerenchuk2018rankdcg} and their RankDCG measure. The main advantage of the RankDCG algorithm is that it provides a clear lower and upper bound for the rank-ordering type of problems. RankDCG is also designed to work with right-skewed data labels, as in our case, where most users have a low k-index, and very few users have a high k-index. RankDCG algorithm scores high k-index users much higher compared to low-rank users. 
Throughout our experiments, we provide the results of four measures: MAE, MSE, nDCG, and RankDCG.

\section{Methods}
\label{sec:methods}

\subsection{Overview}
This section describes our methods for predicting user k-index from a comment. First, we show that the problem is feasible by training a BERT model \cite{devlin2019bert} and using this BERT model as the baseline. Secondly, the BERT model is fine-tuned to achieve the best performance. Third, we train additional seven models for each sub-tasks and use these models to annotate our data with pseudo labels for hedge, age, gender, and four dimensions of MBTI personality types. Lastly, according to our hypothesis, we design a multi-task model that leverages user demographics and personality traits to improve the model's latent representation and achieve better user status prediction results across eight domains.

\subsection{User Status Prediction}
\label{sec:base_bert}
Attention-based models have shown great success on NLP tasks. In this work, we use the small BERT model\footnote{\url{https://tfhub.dev/tensorflow/small_bert/bert_en_uncased_L-4_H-512_A-8/2}} \cite{turc2019} for the following reasons: 1) while there are many transformer-based models, they show only incremental improvements compared to the original BERT model \cite{narang2021transformer}, 2) transformer-based models have high VRAM requirement, which makes them cost-prohibitive in experimental settings. The small BERT allows us to train a model with a batch size of 32 on consumer-grade GPUs within a reasonable time. This small BERT implementation consists of 4 hidden layers of 512 dimensions, each with eight attention heads. Hence, our experiments use the pre-trained small BERT as a base layer.

\begin{figure}
\centering
\includegraphics[width=0.6\linewidth]{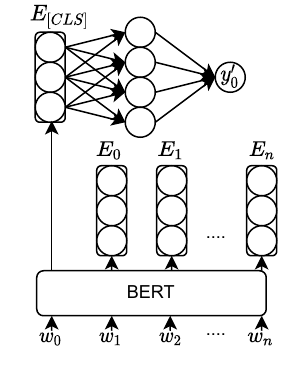}
\caption{Model architecture}
\label{fig:model_architecture}
\end{figure}

The BERT model outputs embedding vectors for the entire text input ($E_{[CLS]}$), and for each token ($E_{1},..,E_{n}$); however, we ignore the word token vectors and only use the input embedding vector ($E_{[CLS]}$). This embedding vector is used as an input to a feed-forward task-specific layer with 256 and an output layer of 1 dimension. Each layer is preceded by a batch normalization layer and a dropout layer set to 0.1. The final output is a linear regression function. In this way, the task is defined as a regression problem where, given a text input, the model predicts the user k-index. The model architecture is shown in Figure  \ref{fig:model_architecture}. 

During the training, the model gradients are calculated with Adam optimizer \cite{kingma2015adam} with a learning rate 5e-5. The objective of the optimization is to minimize mean squared error (MSE). The MSE loss should be more sensitive to large k-index values and "focus" on predicting high-status users. 
The model is trained with an early stopping algorithm that stops training if the MSE of the dev dataset does not change for two epochs. 

In addition to training a BERT model with task-specific layers described above, we train a two-stage fine-tuned model. One of the biggest problems of training a pre-trained model is a phenomenon known as catastrophic forgetting \cite{mccloskey1989catastrophic}. This problem occurs when a model is trained for one task and later re-trained for another. The initial high gradient propagation destroys the learned weights of the first task. Following the steps to mitigate this problem described in \citet{howard2018universal}, we train the model in two stages: 1) freeze the BERT layers and only train the task-specific feed-forward layer. 2) after the first step converges to a local minimum, unfreeze the BERT layers and train the entire model with a smaller learning rate (5e-6). This two-stage fine-tuned model setup is evaluated and the results are reported in Sec. \ref{sec:results}.

\subsection{Pseudo Label Generation}
\label{sec:pseudo_labels}
Our hypothesis states that k-index prediction can be improved by leveraging user-centric information. After reviewing related literature, user characteristics such as age, gender, and use of hedges are associated with user status and status manifestation \cite{rosenthal2014detecting, prabhakaran2014gender, gilbert2012phrases}. In addition, we explore the Myers-Briggs Type Indicator (MBTI) \citep{myers1987description} model associated with user personality traits. These traits correspond to different user behaviors online \cite{wu2017online}. However, our dataset does not include user-related annotations. For this reason, we train separate models to annotate the data with predicted pseudo labels.

\subsubsection{Demographic and Personality Trait Annotations}

To create annotations for age, gender, MBTI types, and hedges, we train a separate model for each task. The training data comes from the PANDORA corpus \cite{gjurkovic2021pandora}. This corpus contains a collection of Reddit users, comments, and labels for age, gender, Extraverted/Introverted, Sensing/Intuitive, Thinking/Feeling, and Judging/Perceiving. Each model is trained to predict users' age (regression) or gender (raw sigmoid score) and MBTI types as a binary label (classification) from a single comment. We sample and balance the data for each task.  While developing the SOTA models for the sub-tasks is not the objective of this paper, we do experiment with different neural network architectures such as LSTM (2 layers, tanh activation), CNN (3 conv. layers with 2, 4, 8 kernel size, relu activation), and the BERT model architecture described in Sec. \ref{sec:base_bert}. Table \ref{table:user_trait_models} reports the best-performing architectures and their scores. 
These models generate pseudo labels for our dataset described in Sec. \ref{sec:reddit}. 

\begin{table}[!th]
\centering
\begin{tabular}{|l|c|r|}
\hline
Task  &  Model  &  Score \\
\hline
Age &  CNN  &  5.81 MAE \\
Gender &  BERT  &  69.60\% ACC \\
Introvert &  CNN  &  56.57\% ACC \\
Intuitive &  BERT  &  54.02\% ACC \\
Perceiving &  BERT  &  52.62\% ACC \\
Thinking &  CNN  &  59.83\% ACC \\

\hline
\end{tabular}
\caption{User Characteristics Sub-task Results}
\label{table:user_trait_models}
\end{table}

\begin{table*}[]
\centering
\begin{tabular}{|l|lllllll|l|}
\hline
 & Age & Women & Introvert & Intuitive & Perceiving & Thinking & Hedge \\
\hline
AskMen       & 28.22 & 0.49 & 0.50 & 0.30 & 0.48 & 0.46 & 0.44 \\
AskScience   & 27.46 & 0.16 & \bf{0.80} & \bf{0.87} & 0.58 & \bf{0.90} & 0.50 \\
AskWomen     & 27.97 & \bf{0.80} & 0.52 & \underline{0.26} & 0.40 & \underline{0.28} & 0.38  \\
Atheism      & 27.56 & 0.25 & 0.75 & 0.55 & 0.60 & 0.74 & 0.51 \\
ChangeMyView & 27.59 & 0.30 & 0.61 & 0.61 & 0.56 & 0.83 & \bf{0.58} \\
Fitness      & \underline{27.02} & 0.36 & \underline{0.46} & 0.42 & \underline{0.29} & 0.63 & \underline{0.36} \\
Politicts     & \bf{28.58} & 0.20 & 0.64 & 0.56 & 0.58 & 0.88 & 0.45 \\
WorldNews    & 27.18 & \underline{0.12} & 0.66 & 0.66 & \bf{0.62} & 0.82 & 0.46 \\
\hline
\end{tabular}
\caption{Predicted pseudo label distributions for each subreddit with max values in bold and min values underscored.}
\label{table:pesudo_label_predictions}
\end{table*}

\subsubsection{Hedge Annotations}
Hedges are linguistic devices commonly used to mitigate orders, statements, or opinions \cite{lakoff1975hedges}. High-status individuals often use hedges to mitigate a statement or an order. Hence, detecting hedge comments can improve k-index prediction. After reviewing recent research in this domain, we use the model proposed by \citep{katerenchuk2021you}. The model is a dual input model of text and part-of-speech (POS) tags. The inputs are fed into two parallel models of LSTM layers for POS and GRU layers for sentences. The latent representation of the LSTM and GRU layers merged into a single layer used as an input to the feed-forward output layer. We choose this model for the following reasons: 1) the model architecture is straightforward, 2) it is efficient regarding training time, and 3) it produces near SOTA results on hedge detection tasks. Their work uses the CoNLL 2010 Wikipedia dataset \cite{farkas2010conll} with binary hedge labels. The dataset contains 11,110 training sentences. Some sentences are short, containing only a few words, such as titles. We ignore those for our problem. After cleaning the data, the training set contains 8,925 data points. After training the model on the pre-processed dataset, the model achieves an F1 score of 67.2\%. We run this hedge detection model to generate pseudo labels on our dataset.

\subsubsection{Pseudo Label Analysis}
\label{sec:pesudo_label_analysis}
Pseudo-label annotation is an excellent way to generate missing labels. However, it is difficult to assess their quality without the actual labels. This brings a question: How accurate is our data annotation? One way to assess the quality is to look at the prediction distributions. In Table \ref{table:pesudo_label_predictions}, we show the mean predictions of each pseudo label with respect to the subreddit. 
While it is hard to interpret all values, 
we can look into the max (in bold) and the min (underscored) values. The table highlights a couple of interesting patterns: the predicted age is the highest in the Politics subreddit and the lowest in Fitness. The assumption is that politics attracts older users, and discussions about fitness for younger users are reasonable. The gender prediction shows that 80\% of female-written comments are in the AskWomen subreddit. This confirms the subreddit's purpose that men ask women questions and, as a result, most answers come from women.
Furthermore, the highest predictions for Introversion, iNtuition, and Thinking (INT) are for the AskScience subreddit. According to the MBTI personality type system, the Scientist (INTJ) is defined with these three dimensions.
Hedge words are the highest in the ChangeMyView subreddit, which aligns with the theory that hedge phrases are used to mitigate statements, sound polite, and influence others' opinions \cite{lakoff1975hedges}. The pseudo labels 
confirm common beliefs in this area.

\begin{figure*}[h]
\centering

\begin{subfigure}{.47\textwidth}
  \centering
  \includegraphics[width=1\linewidth]{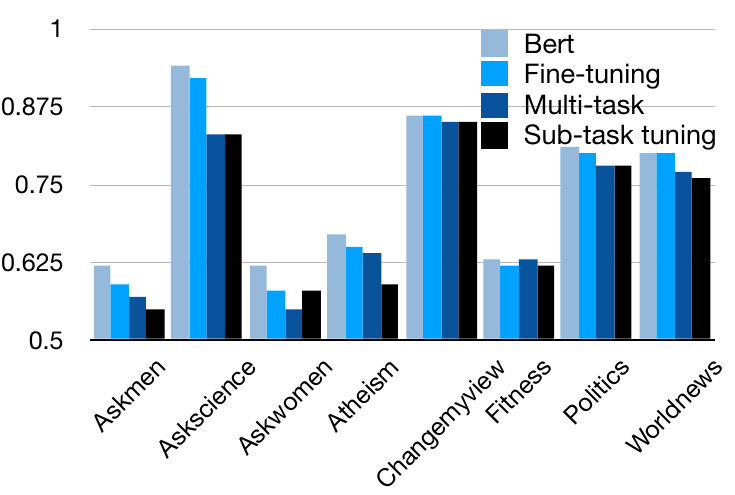}
  \caption{Mean Absolute Error}
\end{subfigure}%
\begin{subfigure}{.47\textwidth}
  \centering
  \includegraphics[width=1\linewidth]{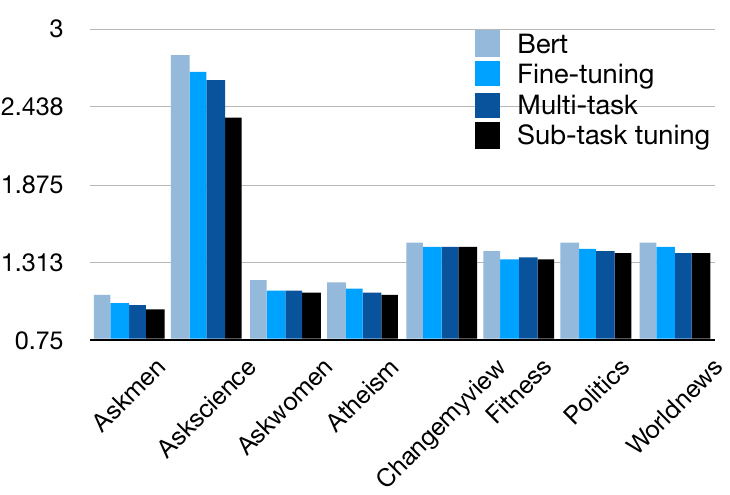}

  \caption{Mean Squared Error}
\end{subfigure}
\begin{subfigure}{.47\textwidth}
  \centering
  \includegraphics[width=1\linewidth]{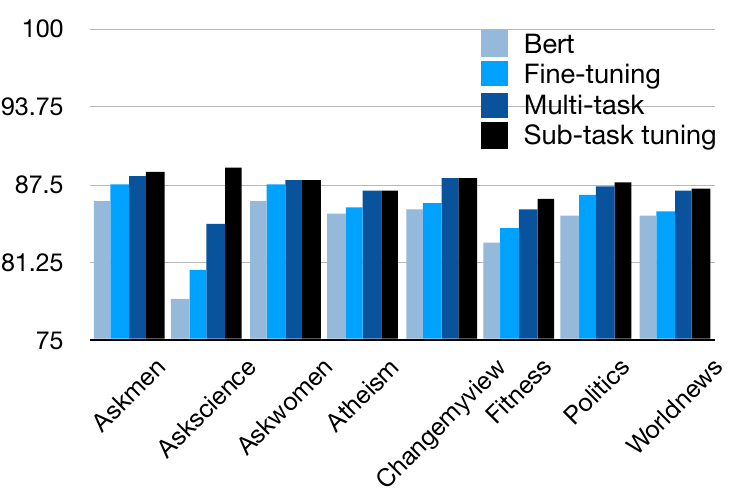}
  \caption{nDCG}
\end{subfigure}%
\begin{subfigure}{.47\textwidth}
  \centering
  \includegraphics[width=1\linewidth]{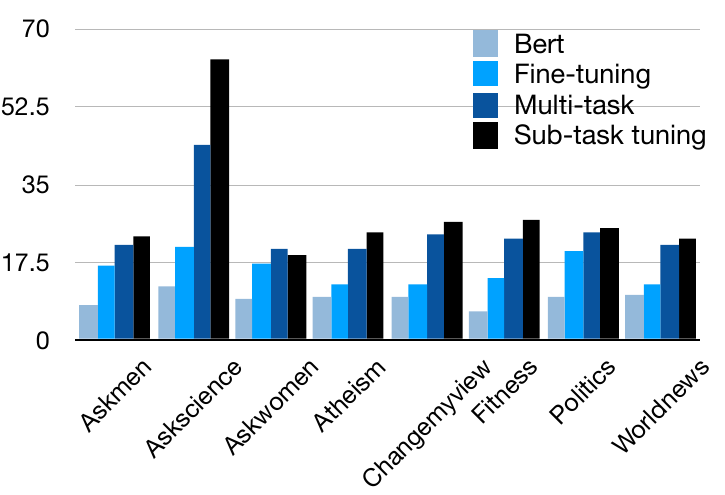}
  \caption{RankDCG}
\end{subfigure}

\caption{Results Across the Subreddits}
\label{fig:results}
\end{figure*}

\subsection{Multi-Task}
\label{sec:multitask}
After generating pseudo labels, we can use this user information by introducing additional learning objectives to improve latent representation. Multi-task learning is an excellent way to introduce additional learning objectives to the model. The multi-task model is an extension of the architecture introduced in \ref{sec:base_bert}. We use the same BERT model and two-layer architecture for the k-index prediction for multi-task learning architecture. However, we add the same two-layer head to the base BERT for each sub-task. In other words, we add seven additional heads with feed-forward layers of 256 and 1 dimension for each sub-task. The output activation function is linear (age) or sigmoid (gender, hedge, MBTI types). 

One drawback of Multi-task learning is that the gradients from sub-tasks can introduce noise, especially when losses have different magnitudes (Ex. MAE vs. MSE). For this reason, we explore tuning loss weights for each task. 

\subsection{Loss Weight Tuning}
\label{sec:loss_weights}
Weight loss tuning for each sub-task is often time-prohibitive. Keras tuner \cite{omalley2019kerastuner} automates this process with search algorithms. Hyperband \cite{li2017hyperband} search algorithm is used in our work to find the best weights. The search algorithm is limited to 20 epochs with early stopping and one iteration. The search space is from 0.0 to 1.0 with a step of 0.1 for each of $w_1, w_2 \ldots w_7$. The total model loss is defined as follows:
\begin{equation}
L_{total} = L_0 + w_1 * L_1 + w_2 * L_2 + \cdots + w_7 * L_7 
\end{equation}
where $L_0$ is k-index loss, $L_i$ - sub-task specific loss, and $w_i$ - the loss weigh. 


\section{Results}
\label{sec:results}
This paper is based on the hypothesis that additional user-centric information can improve user status prediction. To show that the hypothesis stands, we create four models for the k-index prediction task: 1) BERT model, 2) fine-tune BERT, 3) multi-task BERT with user-centric sub-task, and 4) multi-task BERT with tuned loss weights for sub-tasks. The results in Figure \ref{fig:results} are reported as a mean of five runs across four measures: a) MAE, b) MSE, c) nDCG, and d)RankDCG. RankDCG provides a clear measure of our models' performance, with lower and upper bounds being between 0.0 and 1.0 and the emphasis on identifying high-rank users. The figure shows that the base BERT model trained on the k-index prediction task 
achieves 9.51\% RankDCG on average across all subreddits. For example, a zero-shot model (without any training) produces a 0.05\% RankDCG score. By fine-tuning the model, the RankDCG score averaged 15.87\%. These results are achieved by using a user's comment alone. The multi-task BERT model outperformed the previous two architectures by leveraging user-centric pseudo labels, producing the mean RankDCG score of 24.85\%. This improvement shows that the sub-tasks introduce an additional signal that improves k-index prediction. However, multi-task architecture assumes that each sub-task contributes equally to the problem, which might not be the case. For this reason, the last experiment searches through the loss weights for each sub-task to find optimal values. This step further improves the mean RankDCG score to 28.97\%. However, while the average score across all subreddits is higher, the AskWomen subreddit showed lower results when tuning for sub-task loss weights. On the other hand, the AskScience subreddit produces much greater improvements. We believe this could be due to 1) a smaller dataset size (33k vs. 100) and 2) domain-specific language and user behavior. Overall, the results show that introducing additional sub-tasks of user-related information improves the results across eight domains.

\section{Analysis}
This section reflects on our findings by examining our hypothesis and the models. First, we look into sub-task contributions to determine which sub-problems are more salient. Then, we evaluate the latent space. Lastly, we ask questions about whether this approach can improve cross-domain performance.

\subsection{Sub-task Impact}
\begin{figure}
\centering
\includegraphics[width=1\linewidth]{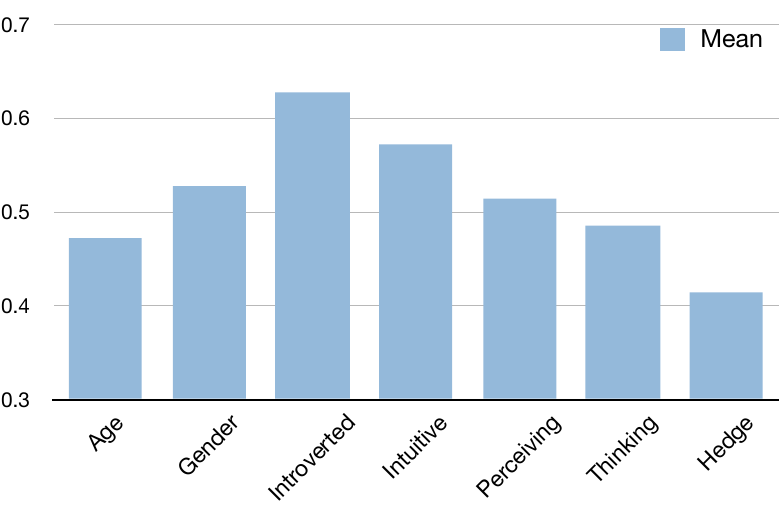}
\caption{Mean Loss Weights}
\label{fig:mean_loss_weights}
\end{figure}

To examine which sub-task is the most impactful, we look at each sub-task weight from sec. \ref{sec:loss_weights}.  
The weights are the results of Hyperband search algorithms that iteratively try different values that lead to the highest result. 
Figure \ref{fig:mean_loss_weights} shows mean loss weight values across all subreddits. The chart shows that Introversion, Intuitive, and Gender sub-tasks are the most prominent and have the highest loss weights. However, by examining loss weights in each subreddit, we can observe that the weights are domain-specific. In the table \ref{table:loss_weights}, we can see that in the AskMen and the AskWomen, gender-centric subreddits, the sub-task for predicting a user's gender plays an important role. The AskScience subreddit relies on Introversion and Thinking prediction sub-tasks. The fitness subreddit also puts weight on 0.9 on the Introversion/Extroversion sub-task. We hypothesize that this is due to extroverts being more active in the subreddit, as was shown in \ref{sec:pesudo_label_analysis}. Another interesting observation can be made on the ChangeMyView subreddit. The highest weights are for the Thinking and Hedge sub-tasks. Such high weights can be a result of subreddit-specific user behavior where users try to influence someone's opinion and use more cognitive effort and hedges in their comments.

\begin{table*}[]
\centering
\begin{tabular}{|llllllll|}
\hline
\textbf{} & Age & Gender & Introverted  & Intuitive & Perceiving & Thinking & Hedge \\ \hline
Askmen  & \textbf{0.8} & \textbf{0.8} & 0.1 & 0.3 & 0.5 & 0.1 & 0.5 \\ 
Askscience & 0.5 & 0.1 & \textbf{0.8} & 0.4 & 0.6 & \textbf{0.8} & 0.6 \\ 
AskWomen & 0.4 & \textbf{0.9} & 0.0 & \textbf{0.7} & 0.3 & 0.3 & 0.0 \\
Atheism & 0.7 & 0.3 & \textbf{0.8} & \textbf{0.8} & 0.4 & 0.3 & \textbf{0.8} \\ 
Changemyview & 0.1 & 0.1 & 0.6 & 0.1 & 0.2 & \textbf{0.9} & \textbf{0.7} \\
Fitness & 0.1 & 0.4 & \textbf{0.9} & 0.5 & \textbf{0.8} & 0.4          & 0.2\\
Politics & 0.2 & \textbf{0.6} & 0.5 & \textbf{0.6} & \textbf{0.7} & 0.4 & 0.1 \\
Worldnews & 0.5 & 0.5 & \textbf{0.7} & \textbf{0.6} & 0.1 & 0.2 & 0 \\ \hline
\end{tabular}
\caption{Loss weights for each sub-task across eight subreddits.}
\label{table:loss_weights}
\end{table*}

\subsection{Latent Space}
\begin{figure}
\centering
\includegraphics[width=1\linewidth]{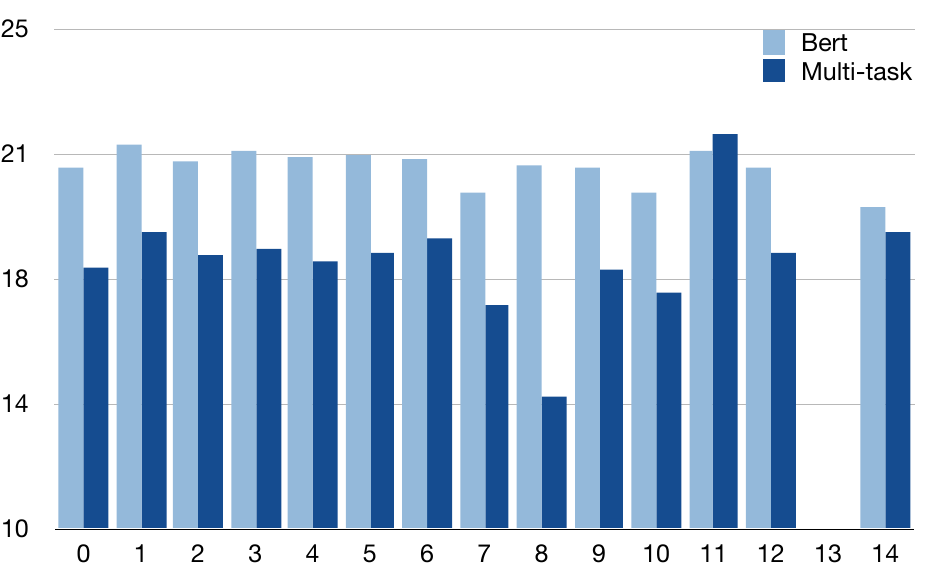}
\caption{Inter-class distances}
\label{fig:inter_class_distances}
\end{figure}

We hypothesize that by introducing additional user-related data, the models leverage this information to improve latent representation, which is beneficial for the task of k-index prediction. To verify this claim, we examine whether we can observe any change in the internal representation. For this reason, we create a PCA projection of a k-index-specific layer of 256 dimensions onto 2-dimensional space. The projections are shown in the Appendix Sec. Figure \ref{fig:latent_space_pca}. While the projections do not show obvious class separation, we can observe that points become more polarized.
Another way to see if the sub-tasks improve latent representation is to calculate inter-class distances. In other words, we calculate the mean of each point to every other point of the same class. 
The smaller distances in Figure \ref{fig:inter_class_distances} show that the same class data points are closer in the latent space.


\subsection{Cross-domain Results}
The sections above show that introducing user-related sub-tasks improves the model's latent representation and performance. In this section, we would like to consider this: Can this approach improve the performance across domains? In NLP, many problems rely on domain-specific language, and the results across domains suffer. We hypothesize that we can improve the performance across different domains by leveraging sub-problems. The idea is that the model would learn universal user representation and use this knowledge to produce more accurate results in a new domain.

We experimented with BERT, fine-tuned BERT, and multi-task BERT models. The task is designed to sample 20k instances from all eight subreddits and use the seven subreddits (140k samples) as a training set and one subreddit as a dev and test set (10k samples each). We repeat this experiment for each subreddit in a leave-one-out fashion.   

After running the experiments, we saw the results of BERT and fine-tuned BERT produced slightly lower scores of 11.6\% and 13.21\% as the mean across all subreddits compared to the results described in sec. \ref{sec:results}. However, adding a user-related sub-task yielded the lowest results, with a mean RankDCG of 5.51\%. Such low performance can be observed across all subreddits. We believe that such low results are due to user differences across each subreddit, which we observe in both tables \ref{table:pesudo_label_predictions} and \ref{table:loss_weights}. The detailed scores are provided in the appendix section in Table \ref{table:cross_domain_results}.

\section{Conclusion}
\label{sec:comclusion}
This work proposes a new method to improve user status prediction on social media. While the area of NLP and user status prediction relies on more extensive and more demanding models, we leverage the earlier work that shows correlations between user traits, demographics, and user status. In particular, we train separate models and use them to annotate our data with age, gender, hedge, and MBTI personality types. All these auxiliary pseudo labels are used in a multi-task BERT model. This approach improves performance over BERT and fine-tuned BERT models. Furthermore, after analyzing the modes, the loss weights support common beliefs in behavior.   

\newpage

\section*{Limitations}
This paper demonstrates that user-centric information can be used to improve user influence level prediction. However, we also discover some limitations of this work.
First of all, this work focuses on the Reddit dataset. While the Reddit website attracts diverse users to discuss various topics, we would like to extend our experiments to different social media platforms to validate our results in future work.   

Second, this work shows a significant improvement on in-domain data; however, it does not generalize well across different domains. After our experiments, we believe that the user behavior is domain-specific. By introducing user-centric sub-tasks, the model learns user characteristics that are irrelevant to this different domain. Such results can be observed in AskScience and Fitness subreddits (Appendix Table \ref{table:cross_domain_results}). These subreddits attract users that generally represent specific traits and behaviors.

Lastly, this work relies on user-centric information to improve influence level prediction. However, we do not have the ground truth labels; the data is annotated with pseudo labels. These auxiliary pseudo-labels provide only partial information about the users. Despite improving results across our eight domains, having the ground truth labels can bring even more significant improvements.

\section*{Ethics Statement}


This research introduces novel methodologies to improve the prediction of user influence levels on social media platforms by incorporating insights from socio-linguistic research, behavioral sciences, and computational models. We use a concept of "influence" that is inherently context-dependent and varies across different domains and settings. Our definition of influence as a function of community endorsement is specific to online social networks and defined as a normalized number of karma points. This work should not be indiscriminately applied to other contexts, such as professional environments, where such algorithms might reinforce existing biases or create new forms of discrimination.

Furthermore, using text classification algorithms and user-centric information raises concerns regarding privacy and potential bias. Models trained on social media data may learn biases present in the input data, associating specific demographics, personality traits, or language use with specific levels of influence. This can lead to unfair characterizations of individuals based on incomplete or biased data sets. For this reason, researchers and practitioners should be aware of model limitations and define transparent guidelines for data. This should include transparency, data anonymization, continuous bias mitigation (by making the model more general with the inclusion of other sources), and prohibiting applications that could harm individuals or society, such as surveillance, profiling, or manipulation.  

In conclusion, while our work contributes valuable insights into predicting user influence levels, these technologies must be developed and used responsibly, with a keen awareness of their ethical implications.

\bibliography{custom,anthology}

\appendix

\section{Example Appendix}
\label{sec:appendix}
\subsection{Influence Level Prediction Tables}
\label{apendix:influence_level_prediction}

\begin{table*}
\begin{tabular}{|l|lllllllll|}
\hline
& Askmen & AskSci & Askwomen & Atheism & CMV & Fit. & Polit. & W.News & Mean \\
\hline
Bert            & 0.62   & 0.94       & 0.62     & 0.67    & 0.86         & 0.63    & 0.81     & 0.80      & 0.66 \\
FT     & 0.59   & 0.92       & 0.58     & 0.65    & 0.86         & 0.62    & 0.80     & 0.80      & 0.65 \\
MT      & 0.57   & 0.83       & 0.55     & 0.64    & 0.85         & 0.63    & 0.78     & 0.77      & 0.62 \\
LWT & 0.55   & 0.83       & 0.58     & 0.59    & 0.85         & 0.62    & 0.78     & 0.76      & 0.62 \\
\hline
\end{tabular}
\caption{MAE results across eight subreddits for Bert, Fine-Tuned Bert (FT), Multi-Task Bert (MT), and Loss Weight Tuned Bert (LWT).}
\end{table*}

\begin{table*}[]
\begin{tabular}{|l|lllllllll|}
\hline
& Askmen & AskSci & Askwomen & Atheism & CMV & Fit. & Polit. & W.News & Mean \\
\hline
Bert            & 1.08   & 2.81       & 1.18     & 1.17    & 1.46         & 1.40    & 1.46     & 1.45      & 1.50 \\
FT     & 1.02   & 2.68       & 1.11     & 1.12    & 1.43         & 1.34    & 1.41     & 1.42      & 1.44 \\
MT      & 1.01   & 2.62       & 1.11     & 1.10    & 1.43         & 1.35    & 1.40     & 1.38      & 1.43 \\
LWT & 0.98   & 2.36       & 1.09     & 1.08    & 1.42         & 1.33    & 1.38     & 1.38      & 1.38 \\
\hline
\end{tabular}
\caption{MSE results across eight subreddits for Bert, Fine-Tuned Bert (FT), Multi-Task Bert (MT), and Loss Weight Tuned Bert (LWT).}
\end{table*}

\begin{table*}[]
\begin{tabular}{|l|lllllllll|}
\hline
& Askmen & AskSci & Askwomen & Atheism & CMV & Fit. & Polit. & W.News & Mean \\
\hline
Bert            & 86.15  & 78.26      & 86.15    & 85.20   & 85.55        & 82.87   & 84.93    & 85.05     & 84.27 \\
FT     & 87.55  & 80.63      & 87.55    & 85.74   & 86.02        & 84.09   & 86.63    & 85.38     & 85.45 \\
MT     & 88.12  & 84.42      & 87.80    & 87.01   & 88.06        & 85.50   & 87.32    & 87.04     & 86.91 \\
LWT & 88.44  & 88.79      & 87.80    & 86.95   & 87.99        & 86.31   & 87.75    & 87.17     & 87.65 \\
\hline
\end{tabular}
\caption{nDCG results across eight subreddits for Bert, Fine-Tuned Bert (FT), Multi-Task Bert (MT), and Loss Weight Tuned Bert (LWT).}
\end{table*}

\begin{table*}[]
\begin{tabular}{|l|lllllllll|}
\hline
& Askmen & AskSci & Askwomen & Atheism & CMV & Fit. & Polit. & W.News & Mean \\
\hline
Bert            & 8.15   & 12.21      & 9.41     & 9.91    & 9.82         & 6.41    & 9.94     & 10.24     & 9.51  \\
FT     & 16.73  & 21.12      & 17.13    & 12.46   & 12.83        & 13.95   & 20.00    & 12.71     & 15.87 \\
MT      & 21.32  & 43.74      & 20.73    & 20.62   & 23.80        & 22.91   & 24.31    & 21.36     & 24.85 \\
LWT & 23.46  & 63.00      & 19.09    & 24.42   & 26.65        & 26.87   & 25.35    & 22.89     & 28.97 \\
\hline
\end{tabular}
\caption{RankDCG results across eight subreddits for Bert, Fine-Tuned Bert (FT), Multi-Task Bert (MT), and Loss Weight Tuned Bert (LWT).}
\end{table*}

\begin{table*}[]
\begin{tabular}{|l|llllllll|}
\hline
& Askmen & Askscience & Askwomen & Atheism & CMV & Fitn. & Polit. & Worldnews \\
\hline
Bert        & 8.56   & 8.69       & 10.31    & 16.36   & 16.20        & 9.60    & 12.14    & 10.98     \\
Fine-tuning & 12.08  & 11.41      & 13.57    & 9.36    & 22.02        & 10.22   & 12.81    & 14.26     \\
Multi-task  & 5.70   & 1.94       & 7.37     & 4.26    & 5.67         & 2.39    & 6.51     & 10.24 \\
\hline
\end{tabular}
\caption{Cross-domain RankDCG scores}
\label{table:cross_domain_results}
\end{table*}

\begin{figure*}[th]
\centering
\begin{subfigure}{.48\textwidth}
  \includegraphics[width=1\linewidth]{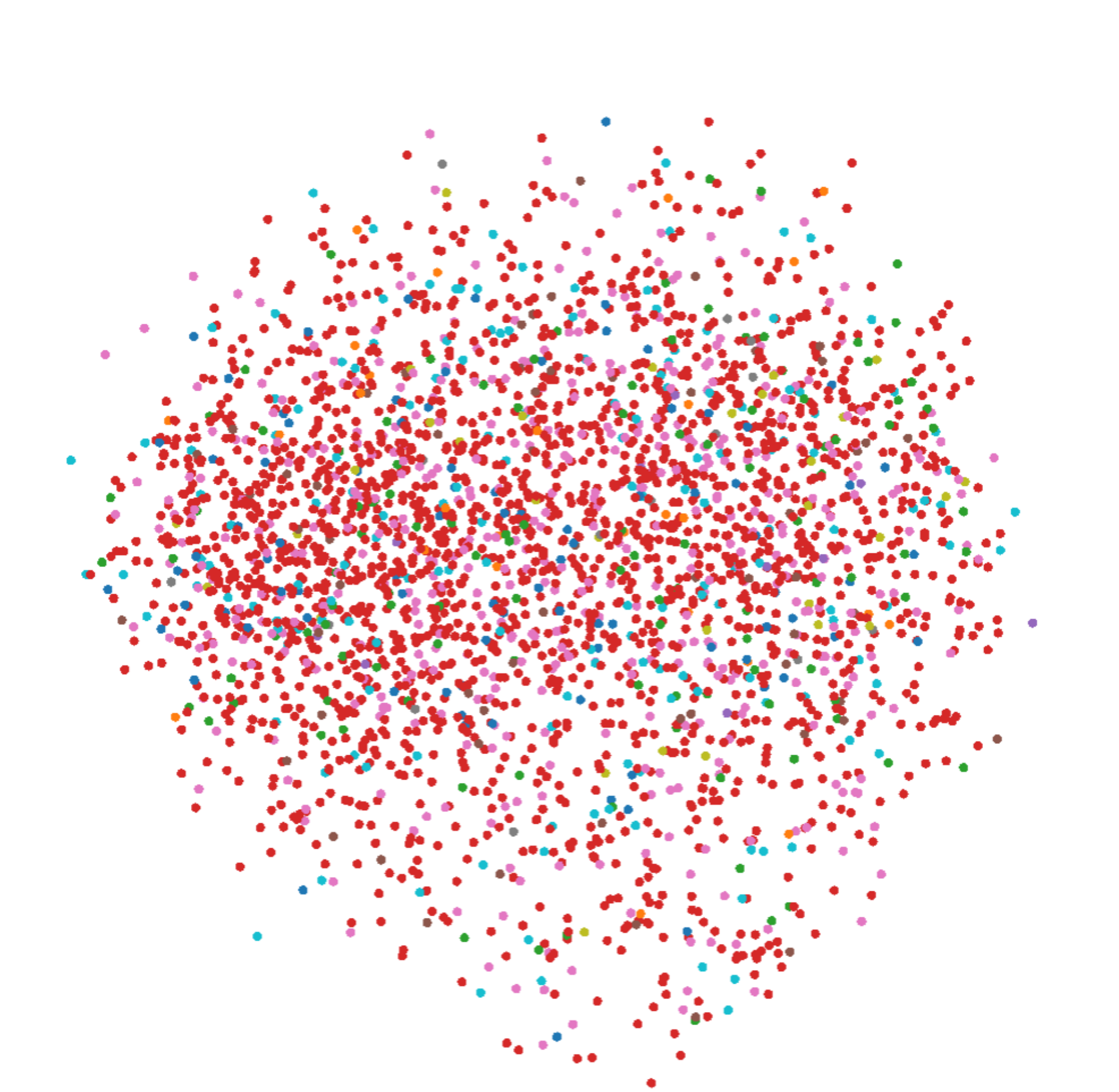}
  \caption{{Latent Space of Bert-based Model }}
\end{subfigure}%
\begin{subfigure}{.48\textwidth}
  \includegraphics[width=1\linewidth]{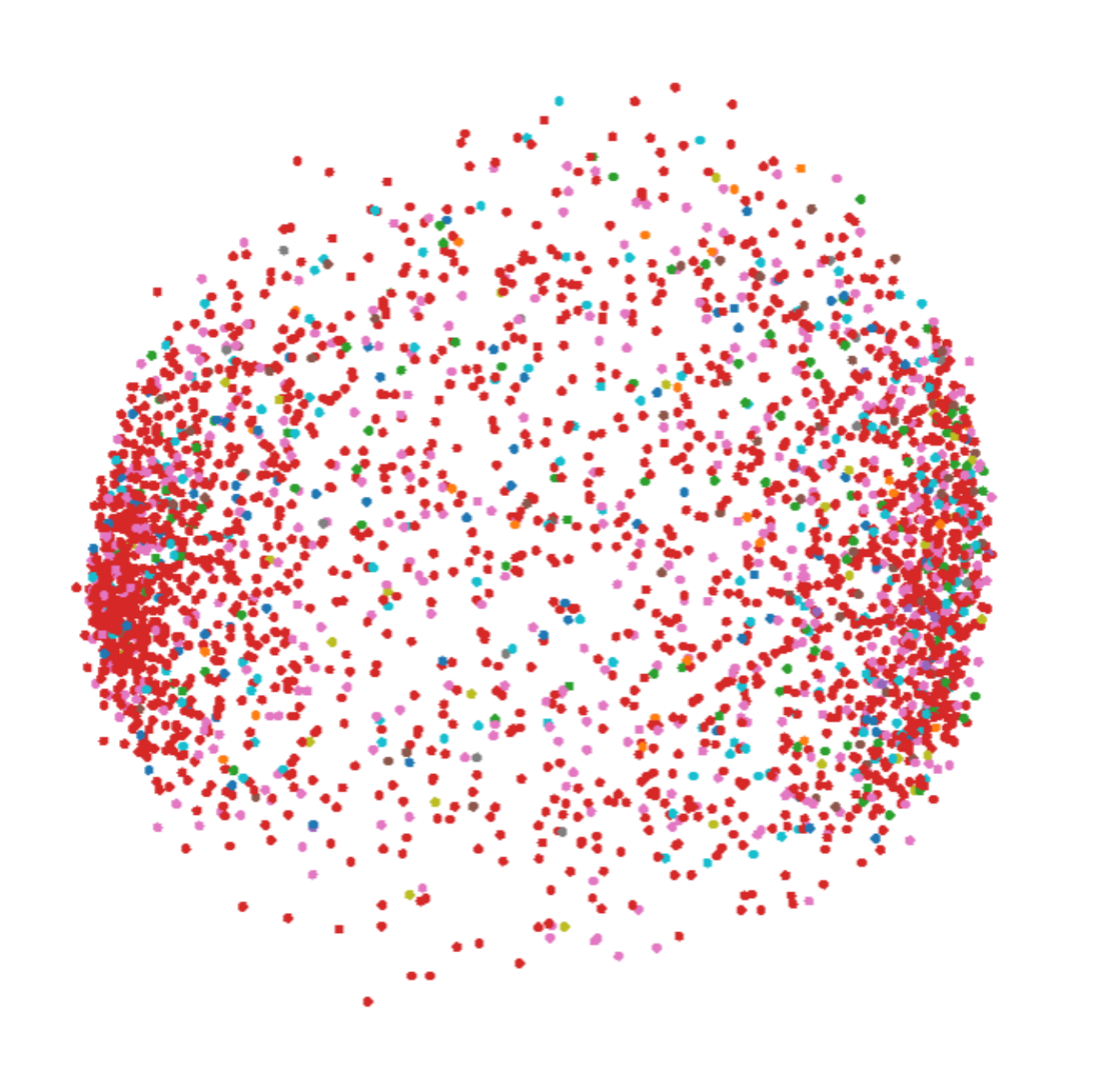}
\caption{Latent Space of Multi-label Model}
\end{subfigure}
\caption{Latent Space PCA projection. While the projection does not show obvious clustering, the latent representation becomes more polarized}
\label{fig:latent_space_pca}
\end{figure*}



\end{document}